# High transport critical current densities in textured Fe-sheathed $Sr_{1-x}K_xFe_2As_2$+Sn superconducting tapes

Zhaoshun Gao, Lei Wang, Chao Yao, Yanpeng Qi, Chunlei Wang, Xianping Zhang, Dongliang Wang, Chengduo Wang, Yanwei Ma*

Key Laboratory of Applied Superconductivity, Institute of Electrical Engineering, Chinese Academy of Sciences, Beijing 100190, China

**Abstract:**

We report the realization of grain alignment in Sn-added $Sr_{1-x}K_xFe_2As_2$ superconducting tapes with Fe sheath prepared by *ex-situ* powder-in-tube method. At 4.2 K, high transport critical current densities $J_c$ of $2.5 \times 10^4$ A/cm$^2$ ($I_c$ = 180 A) in self-field and $3.5 \times 10^3$ A/cm$^2$ ($I_c$ = 25.5 A) in 10 T have been measured. These values are the highest ever reported so far for Fe-based superconducting wires and tapes. We believe the superior $J_c$ in our tape samples are due to well textured grains and strengthened intergrain coupling achieved by Sn addition. Our results demonstrated an encouraging prospect for application of iron based superconductors.

* Author to whom correspondence should be addressed; E-mail: ywma@mail.iee.ac.cn

The discovery of superconductivity in LaFeAsO$_{1-x}$F$_x$ and related compounds, with a higher transition temperature, $T_c$ of ~55 K, has triggered great research interests from both theoretical and applied aspects [1-7]. In addition to $T_c$, these iron based superconductors were reported to have a very high upper critical field, $B_{c2}$, and low $B_{c2}$ anisotropy, making them potential candidates for a wide array of future applications [8–11]. The early results indicate that the global critical current is limited by intergrain currents over the grain boundaries in polycrystalline bulk and wires [12–18]. Recent experiments revealed that high-angle grain boundary largely deteriorates critical current density $J_c$ in Ba(Fe$_{1-x}$Co$_x$)$_2$As$_2$ bicrystals [19, 20]. These results suggest that the discovered iron based superconductors are exhibiting weak link grain boundary behavior similar to high-$T_c$ cuprate superconductors. An effective method to overcome the weak link problem is to engineering textured grains in iron based superconductors to minimize deterioration of the critical current density across high-angle grain boundaries.

Recently Co-doped Ba122 coated conductors have been grown by several groups utilizing the existing YBCO coated conductor technology and have reached a self-field $J_c$ over 1 MA/cm$^2$ [21-23]. However, the technology has the shortcoming of low production rate, complexity and high equipment cost. Furthermore, it is hard to be applied to the volatile elements in iron based superconductors such as alkali metals doped 122 phase and F doped 1111 phase. Cold deformation process is a well-developed technique used to enhance the degree of grain alignment and critical current density of Bi2223 superconductors [24, 25]. It is unsurprising that this technique may also be suitable for iron based superconductors too. Besides grain alignment, adding metallic elements is another effective way to improve the grain connectivity of the iron based superconductors. For example, we have reported that the superconducting properties of the 122 phase iron based superconductor can be significantly increased by Ag or Pb addition [26, 27]. Recently Togano et al [28] reported further improvement in transport critical current in the Ba122 wires with Ag addition. In this work, we report addition of Sn in Fe sheathed Sr$_{0.6}$K$_{0.4}$Fe$_2$As$_2$ tapes for improved superconducting properties. We observed texture and excellent transport $J_c$ in samples prepared by cold working process. The mechanism for significant improvement in transport $J_c$ will be discussed.

The $Sr_{1-x}K_xFe_2As_2$ polycrystalline precursors were prepared by a one-step PIT method developed by our group [29]. Mixtures of Sr filings, Fe powder, As and K pieces were ground using a ball milling method in Ar atmosphere for ≈15 h. In order to compensate for loss of elements during the milling and sintering procedures, the starting mixture contains 10% excess As and 10-20% excess K. Raw powders were encased and sealed into Nb tubes. Finally, the samples were heat treated at 900°C for 35 h. The precursors were ground to a powder in an agate mortar and pestle. For Sn doped samples, 10 wt% Sn was added to the precursor powder and then the mixture was ground in a mortar for half an hour. The final powders were filled and sealed into an iron tube (OD: 8 mm, ID: 5 mm), which was subsequently swaged and drawn down to a wire of 1.9 mm in diameter. The as-drawn wires were then cold rolled into tapes (0.6 mm in thickness) with a reduction rate of 10~20%. Fe is the suitable sheath material for the application of iron based superconductor. It is a cheap and abundant resource. Furthermore, Fe has enough hardness to induce the grain texture during the mechanical rolling process. To minimize the reaction between the Fe sheath and the superconducting core, the tapes were finally sintered at 1100°C for a short time of 0.5~15 minutes in Ar atmosphere [29].

Phase integrity and texture of $Sr_{0.6}K_{0.4}Fe_2As_2$ grains were investigated using x-ray diffraction (XRD), for which iron sheath was mechanically removed after cutting the edges of the tape. Microstructure was studied using a scanning electron microscopy (SEM). Standard four probe resistances were carried out using a physical property measurement system (PPMS). The transport current $I_c$ at 4.2 K and its magnetic field dependence were evaluated at the High Field Laboratory for Superconducting Materials (HFLSM) at Sendai, by a standard four-probe resistive method, with a criterion of 1 μV/cm, a magnetic field up to 10 T was applied parallel to the tape surface. For each set of doped tapes, $I_c$ measurement was performed on several samples to ensure the reproducibility.

Typical SEM microstructures for fracture surface of $Sr_{0.6}K_{0.4}Fe_2As_2$ tapes are shown in Figure 1. It exhibits a dense layered structure, very similar to what has been observed in Bi2223 superconductors. EDX analysis on a large area in the Sn added samples clearly demonstrates that the sample is composed of Sr, K, Fe, As and Sn elements.

The *c*-axis texture was confirmed by x-ray diffraction analysis revealing strong (00*l*) orientation peaks. X-ray diffraction patterns for $Sr_{0.6}K_{0.4}Fe_2As_2$ precursor bulk, grain aligned pure and Sn added tapes are shown collectively in Figure 2. Unlike randomly orientated $Sr_{0.6}K_{0.4}Fe_2As_2$ precursor, which show a strongest non *ab*-plane reflection of (103). The enhanced (00*l*) peaks for the *c*-axis of the grain were observed in the pure and Sn added tapes, but small peaks associated with the *ab*-plane were also observed. It indicates that the grain alignments are still not perfect in the present samples. Although clear peaks of Sn were observed in Sn added tapes before annealing (not shown in this paper), the peaks of Sn disappeared while peaks for possible impurity phase of KSnAs emerged after heat treatment. Due to the high background of XRD data caused by cold working, it needs further investigation to distinguish the impurity phase.

Figure 3 displays the temperature dependence of the normalized resistance for pure and Sn added tapes respectively. The normal state resistivity of our samples show broad bumps around 170 K, which is ascribed to the SDW transition. It suggests that partial K loss during high temperature annealing. Expanded scale for temperatures near the superconducting transition is shown in inset of Fig. 3. Both samples exhibit a sharp resistive superconducting transition. The onset of superconductivity at 36.5 K is achieved in the pure $Sr_{0.6}K_{0.4}Fe_2As_2$ tapes. For the Sn doped sample, a small depression in $T_c$ was observed. As reported by Ni et al. [30], small amounts of Sn can be incorporated into the structure of $Ba_{1-x}K_xFe_2As_2$ and suppress the $T_c$. Therefore, these results imply that some amounts of Sn were substituted into $Sr_{0.6}K_{0.4}Fe_2As_2$ in our samples.

The current-voltage characteristics at 4.2 K in magnetic fields up to 10 T for Sn added tapes are shown in Figure 4a. For Batch I, the critical currents of ~180 A and ~19 A determined by the 1 μV /cm, were clearly seen for self-field and 10 T, respectively. Figure 4b shows the field dependence of critical current density at 4.2 K for pure and Sn added tapes. We also have measured the transport critical current of $Sr_{0.6}K_{0.4}Fe_2As_2$ round wires prepared use the same condition. However, almost no transport currents were observed the in these samples. As a reference, the critical currents of the round wires reported in our previous paper are also included in the figure [31]. As can be seen from Fig. 4b, all textured tapes show a more than one order of magnitude enhancement in $J_c$ compared to round wires and exhibit a very weak field dependence behavior. It worth noting that further improvement in $J_c$ values were achieved with Sn addition in $Sr_{0.6}K_{0.4}Fe_2As_2$ tapes. The Sn doped tapes reveal the highest $J_c$ compared to all other samples in our experiment: At 4.2 K, the transport $J_c$ in self-field and 10 T exhibited values of $2.5 \times 10^4$ A/cm$^2$ and $3.5 \times 10^3$ A/cm$^2$, respectively, which are the highest values ever reported so far for the iron based superconducting wires and tapes. These $J_c$ values of the samples investigated in this work were almost three times larger than those of silver doped wires [28], and highlight the importance of grain alignment and Sn addition for enhancing the $J_c$ of $AFe_2As_2$(122) superconducting wires and tapes. We should note that the grain alignment is still not perfect in the present sample, suggesting that a further enhancement in the $J_c$ performance can be expected with a higher degree of grain alignment.

As we know, the misalignment in crystalline orientation at grain boundaries is a crucial weak-link in connectivity. The well developed grain texture in our samples is one of reasons for the $J_c$ improvement. However, the aligned grains alone are not enough to achieve the high $J_c$ values as shown in Fig. 4b. Another connectivity limit is related to nonsuperconducting second phases within the grain boundaries [13, 14, 32]. For example, Wang et al. [32] found the grain boundaries in Sr122 are coated by amorphous oxide layers, and concluded that the supercurrent across the grain boundaries can be largely reduced by these amorphous, insulating layers. Sn is a good metallic flux for the growth of 122 phase [30]. During the heat treatment at high temperatures, Sn can promote the crystallization at grain boundaries, diminish the formation of amorphous layer, and hence improve intergranular connectivity.

Furthermore, similar to the case of Ca doping in YBCO, the partial $Sn^{4-}$ substitution for $As^{3-}$ may increase the hole concentration of Sr122 and thus enhance transport $J_c$. There is a need for further research to clarify the role of Sn in Sr122 tapes.

In conclusion, significantly improved in-field $J_c$ values were obtained for the textured $Sr_{1-x}K_xFe_2As_2$ superconducting tapes with Fe sheath prepared by applying a combination of the cold working process and Sn addition. We believed that these improvements should be attributed to (i) well aligned superconducting grains and (ii) strengthened intergrain coupling obtained by Sn addition. The $J_c$ will be further enhanced by optimization of annealing process and introduction of pinning centers.

**Acknowledgements**

The Authors thank Profs. S. Awaji and K. Watanabe at the IMR, Tohoku Univ., Japan for the high field transport measurements. They are also indebted to Dr. Ma at ANL for useful suggestion. This work is partially supported by the National '973' Program (Grant No. 2011CBA00105) and National Natural Science Foundation of China (Grant No. 51002150 and 51025726).

# Captions

Figure 1 Typical microstructures of fracture surfaces of $Sr_{0.6}K_{0.4}Fe_2As_2$ tapes. (a) Cross-sectional view SEM image of pure tape, (b) cross-sectional view SEM of Sn added tape, (c) plan view SEM of pure tape, (d) EDX spectrum of Sn added tape.

Figure 2 XRD patterns of tape surfaces after peeling off the iron sheath. The peaks of Fe were contributed from the Fe sheath.

Figure 3 Temperature dependence of the normalized resistance for pure and Sn added $Sr_{0.6}K_{0.4}Fe_2As_2$ tapes at zero field up to 300 K. The inset shows the expanded scale for temperatures near the superconducting transition.

Figure 4 (a) The current-voltage characteristics at 4.2 K in magnetic fields up to 10 T for Sn added tapes. (b) The field-dependent critical current density $J_c$ at 4.2 K in magnetic fields up to 10 T for pure and Sn added tapes. The data of round wires are also included for comparison.

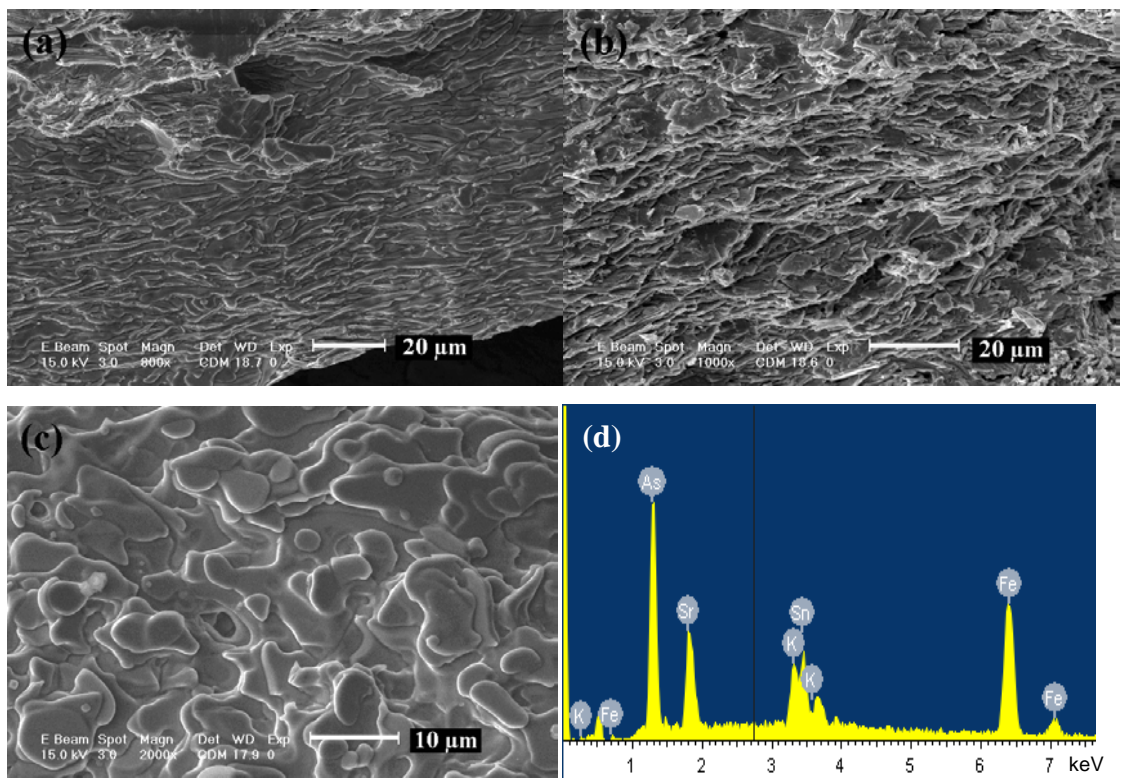

Fig.1 Gao et al.

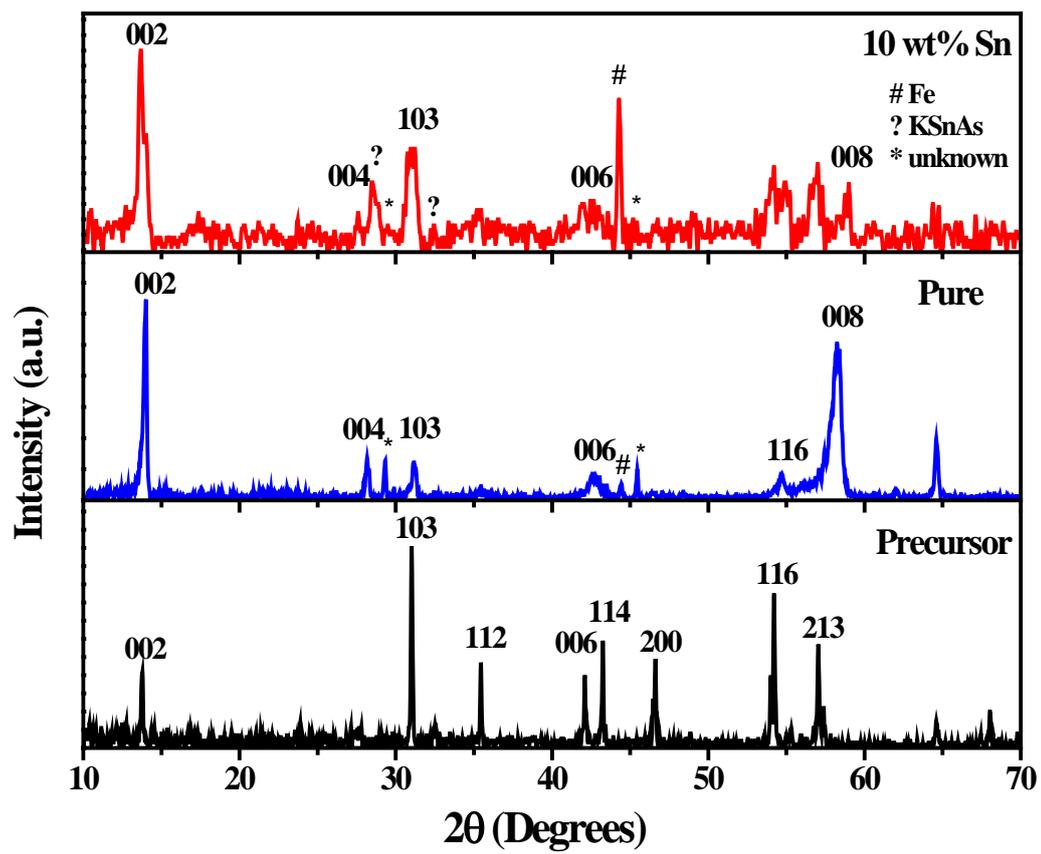

Fig.2 Gao et al.

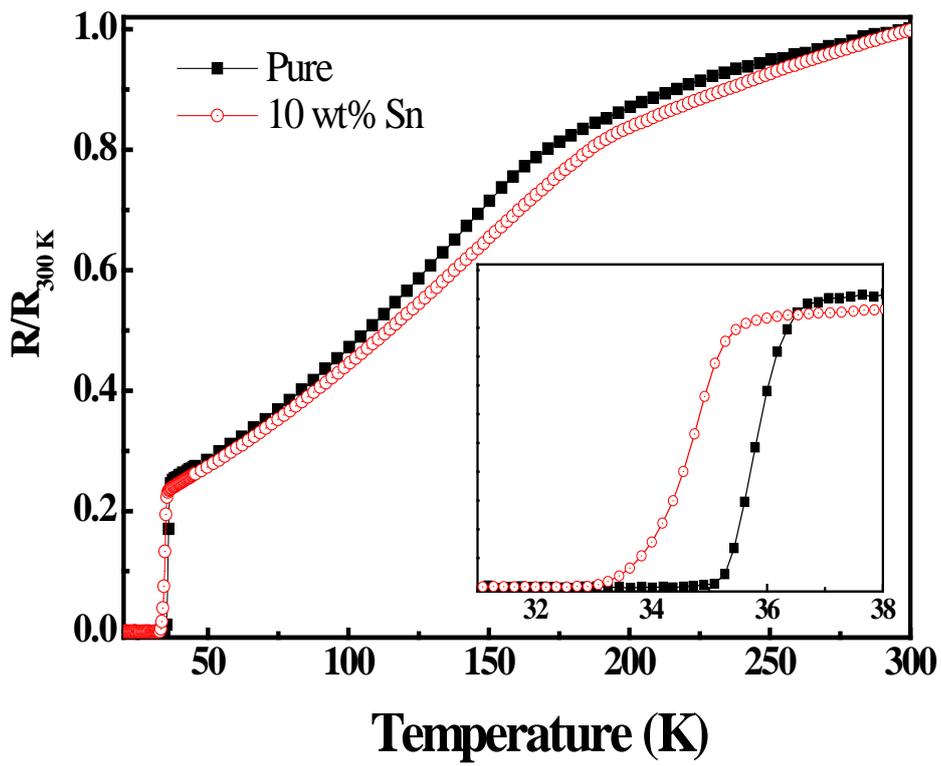

Fig.3 Gao et al.

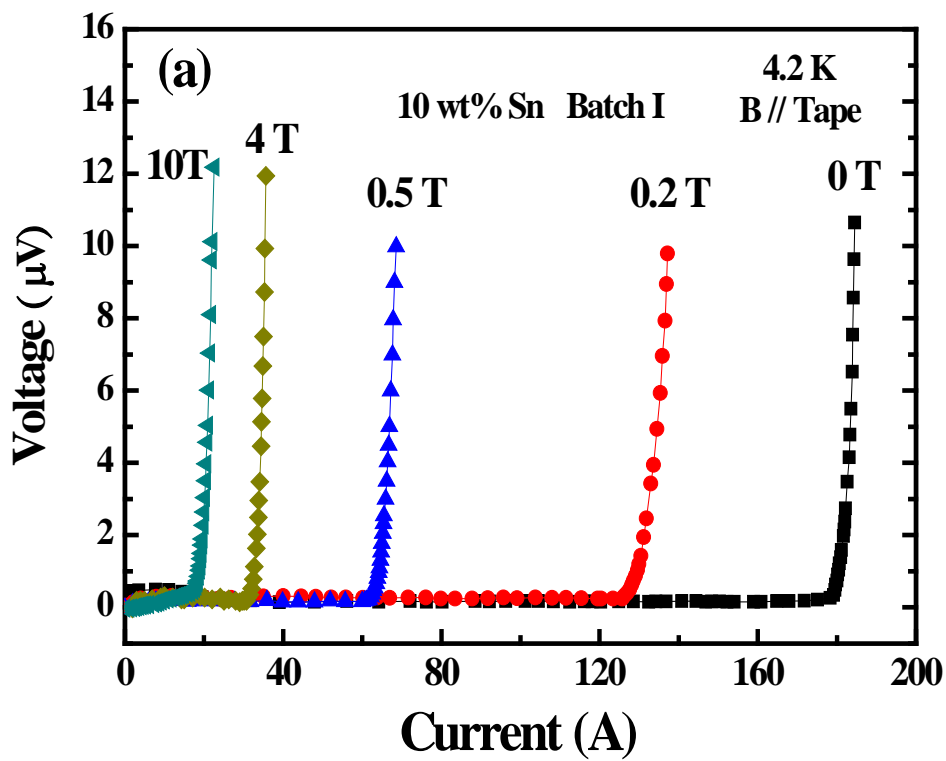

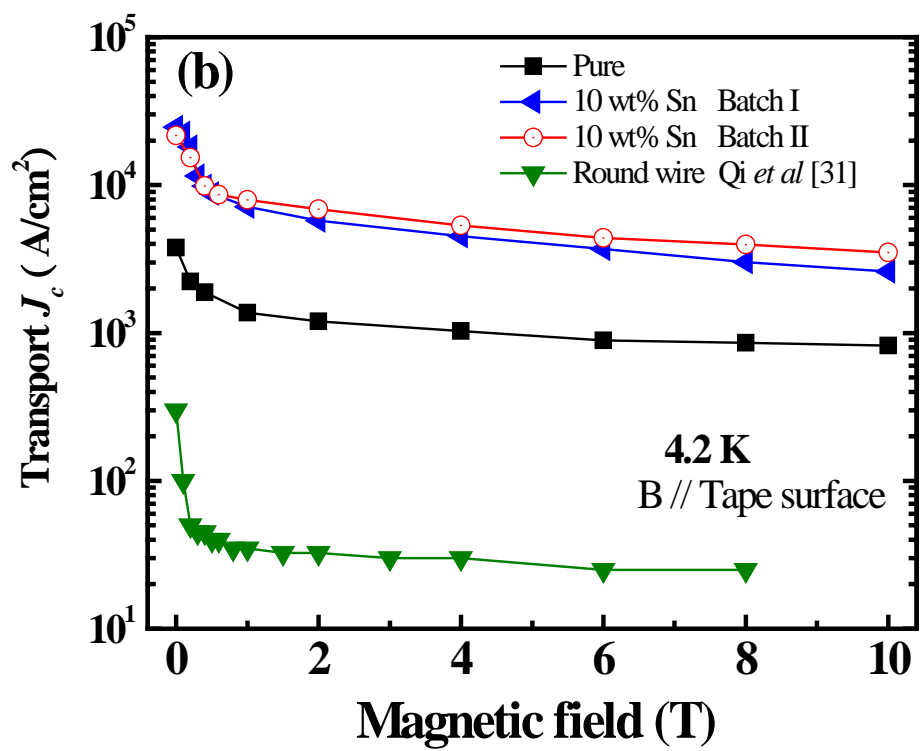

Fig.4 Gao et al.